\documentclass[a4paper,11pt]{article}
\pdfoutput=1 
\usepackage{jheppub} 
\usepackage{hyperref}
\usepackage{booktabs}
\usepackage[T1]{fontenc} 
\usepackage{siunitx} 
\usepackage{graphicx}%
\usepackage[dvipsnames]{xcolor}
\usepackage{dcolumn}
\usepackage{bm}
\usepackage[normalem]{ulem}
\usepackage{braket} 
\usepackage{amsmath}
\usepackage{cancel}
\usepackage{slashed}
\usepackage{multicol}
\usepackage[capitalise]{cleveref}

\counterwithin*{equation}{section}
\allowdisplaybreaks

\newcommand{\nnl}{\nonumber \\}

\newcommand{\cO}{\mathcal{O}}

\newcommand{\cM}{{\mathcal M}}

\newcommand{\re}{{\mathrm{Re}} \,}
\newcommand{\im}{{\mathrm{Im}} \,}

\newcommand{\hc}{\rm h.c.}

\usepackage{etoolbox,siunitx}
\robustify\bfseries

\usepackage{subcaption}

\newcommand{\N}{{\cal N}}
\newcommand{\K}{{\cal K}}

\begin{document}

\title{
On the Coulomb corrections in nuclear beta decay 
}

\author{Edoardo Alviani and}
\author{Adam Falkowski }

\affiliation{Universit\'{e} Paris-Saclay, CNRS/IN2P3, IJCLab, 91405 Orsay, France}

\emailAdd{
alviani@ijclab.in2p3.fr,
adam.falkowski@ijclab.in2p3.fr
}

\abstract{

We propose a Lorentz invariant and little group covariant description of beta decay amplitudes relying on on-shell amplitude methods and the spinor variables for massive particles. 
The framework is employed to calculate Coulomb corrections to the decay amplitude and their contribution to $T$-odd correlation coefficients, including the $D$ parameter. 
In the SM limit we recover the known results for the Coulomb contributions to $D$ and update their numerical values. 
We also calculate new subleading contributions to $D$ in the presence of non-standard scalar and tensor interactions. 
We also point out that another $T$-odd correlation coefficient is generated in the SM at the same order as the $D$ parameter, and provide their numerical values for selected transitions. 
}
\maketitle

\section{Introduction} 
\label{sec:INTRO}

Theoretical description of nuclear beta decay goes back to the original paper of Fermi~\cite{Fermi:1933jpa}. 
For neutron decay, the involved nucleons have spin $1/2$ and the process can be readily described using an effective field theory (EFT) framework rooted in relativistic quantum field theory (QFT), as was already done in the seminal work of Lee and Yang~\cite{Lee:1956qn}. 
However, many interesting beta decay processes involve nuclei with spins $J \geq 1$, where the standard relativistic QFT description becomes awkward, more and more so with each increasing unit of spin. 
This can be addressed by choosing a specific frame, usually the rest frame of the parent nucleus, where nuclear 3-momenta are small compared to their masses. 
This makes the nuclei amenable to non-relativistic EFT description, with expansion in 
$|\boldsymbol{q}|/m_{\cal N} \ll 1$, 
where $\boldsymbol{q}$ is the recoil 3-momentum of the daughter nucleus in parent's rest frame, and $m_{\cal N}$ is the overall mass scale of the involved nuclei.  

In a non-relativistic setting,
handling any spin is straightforward,   following the familiar rules of rotational symmetry.  
Still, this approach may not be optimal beyond the leading order in $|\boldsymbol{q}|/m_{\cal N}$ expansion. 
One reason is that counting independent EFT operators  becomes less transparent in the absence of Lorentz invariance. 
Another is that Lorentz invariance implies non-trivial relations between Wilson coefficients of distinct operators at different orders in the EFT expansion~\cite{Holstein:1974zf}. 

Meanwhile, an efficient formalism to describe massive particles of any spin was proposed several years ago~\cite{Arkani-Hamed:2017jhn}, based on on-shell amplitude methods and the spinor helicity variables. 
The formalism allows one to write down amplitudes for any processes in a manifestly Lorentz invariant way. 
Spins of intervening particles are transparently encoded via transformation properties under the corresponding little group ($U(1)$ and $SU(2)$ for massless and massive particles, respectively).  
The foremost goal of this paper is to adapt this formalism to describe nuclear beta decay. 
We do it explicitly for mixed Fermi--Gamow-Teller allowed transitions (where parent and daughter nuclei have the same spin and belong to the same isospin multiplet), although there are no roadblocks whatsoever to apply it to any other beta processes.  

As an application,  we compute  one-loop Coulomb corrections to the decay amplitudes due to the electromagnetic final-state interactions.   
This can be efficiently done by exploiting unitarity, which implies that discontinuities of one-loop amplitudes (which are directly related to Coulomb corrections) can be expressed as products of tree-level amplitudes integrated over the two-body phase space of intermediate particles. 
In our calculation, Lorentz invariance and little group covariance is manifestly maintained at each step.

Given the Coulomb corrections, we calculate their contributions to the correlation coefficients in the differential decay width, which are observables in nuclear physics experiments.   
We focus on the $T$-odd correlations, especially on the so-called $D$ parameter~\cite{Jackson:1957zz}, which has a potential to probe new sources of CP violation beyond the Standard Model (SM)~\cite{Herczeg:2001vk,Drees:2003dv,Ng:2011ui,El-Menoufi:2016cfo,Ramsey-Musolf:2020ndm,Falkowski:2022ynb}.  
Coulomb corrections contribute to the $D$ parameter in the SM limit~\cite{Callan:1967zz,Chen:1969hkp,Holstein:1972bbl,Ando:2009jk,Cirigliano:2022hob}. 
So far, upper limits on $|D|$ have been placed for neutron~\cite{Chupp:2012ta} and ${}^{19}$Ne~\cite{Hallin:1984mr} decays. 
Future experiments with ${}^{23}$Mg at the DESIR facility at GANIL are expected to reach the sensitivity to detect the $D$ parameter at the level predicted by the SM~\cite{Delahaye:2018kwf,Goyal:2023umy,Tschopp:2023lwb}, providing a direct test of theoretical calculations of radiative effects in beta decay.
The SM calculation of $D$ within our Lorentz invariant framework agrees, when evaluated in the rest frame, with the results in the previous literature~\cite{Callan:1967zz,Holstein:1972bbl}. 
For selected transitions, we provide updated numerical values for $D$  based on the most recent experimental input.  
We also point out that another $T$-odd correlation coefficient is generated in the SM at the same order as the $D$ parameter. 
Potential detectability of this coefficient has not yet been studied. 
Moreover, we also calculate corrections to $D$ due to the possible presence of non-SM scalar and tensor interactions in the beta amplitude.

This paper is organized as follows. 
In \cref{sec:LIBD} we review the spinor helicity formalism for massive particles, and we write down amplitudes for mixed Fermi--Gamow-Teller decays in a Lorentz invariant and little group covariant form. 
We also expand this amplitude in powers of momentum exchange, in which case information about spin and polarization can be condensed into a simple relativistic object - the spin vector. 
In \cref{sec:CC} we calculate the Coulomb corrections in the SM limit by gluing the tree-level beta decay and electromagnetic scattering amplitudes and integrating them over the two-body phase space of intermediate particles. 
The calculation is organized as an expansion in inverse powers of the nuclear mass scale, and includes subleading effects in that expansion. 
The discontinuities calculated in that section contain complete information about Coulomb corrections to all correlations coefficients. 
They are used in \cref{sec:D} to extract the Coulomb corrections to the $D$ parameter and other $T$-odd correlations. 
In \cref{tab:results} we provide updated numerical values of these correlations for selected beta transitions of phenomenological interest. 
Finally, in \cref{sec:NONSM} we calculate Coulomb corrections and their effects on $D$ in the  presence of non-standard scalar and tensor currents. 
Conclusions and future directions are presented in \cref{sec:conclisions}. 
In \cref{app:CON} we provide useful formulae to manipulate the spin vector and leptonic polarization sums, while \cref{app:INT} defines our Lorentz invariant parametrization of the two-body phase space.

\section{Lorentz-invariant formulation of nuclear beta decay amplitudes} 
\label{sec:LIBD}

In this section we introduce a manifestly relativistic formulation of beta decay for arbitrary nuclear spins. 
The construction is based on the spinor-helicity formalism of Ref.~\cite{Arkani-Hamed:2017jhn}. 
This allows us to build quantum amplitudes that are covariant under the little group transformations. 
The spinors transparently encode the spins and polarizations of the massive and massless particles involved in beta decay. 

Let us start by reviewing how to associate spinor variables to massless and massive momenta.\footnote{See in particular Ref.~\cite{Dreiner:2008tw} for conventions and many useful formulae of the spinor algebra.}
First, one defines the Lorentz vectors of sigma matrices: 
$\sigma^\mu = (1, \boldsymbol{\sigma})$, 
$\bar \sigma^\mu = (1, - \boldsymbol{\sigma})$,
with $\boldsymbol{\sigma} \equiv (\sigma^1,\sigma^2,\sigma^3)$ denoting the usual triplet of Pauli matrices.
A massless momentum $p^\mu$, $p^2 =0$, can  be associated with a pair of two-component commuting spinors 
$\lambda_{\alpha}$, 
$\tilde \lambda_{\dot \alpha}$,
$\alpha=1,2$, via the relation 
$p \sigma \equiv 
p^\mu  [\sigma_\mu]_{\alpha \dot \beta} = \lambda_{\alpha} \tilde \lambda_{\dot \beta}$. 
The spinor indices can be raised  by an antisymmetric $\epsilon$ tensor: 
$\lambda^\alpha = \epsilon^{\alpha \beta} \lambda_\beta$, and idem for $\tilde \lambda$.
One can then show that 
$[p \bar \sigma]^{\dot \alpha \beta} 
= \tilde \lambda^{\dot \beta } \lambda^\alpha$, 
and 
$(\lambda  \lambda) \equiv
 \lambda^\alpha  \lambda_\alpha  =0$, 
 $(\tilde \lambda \tilde \lambda) \equiv
 \tilde \lambda_{\dot \alpha} \tilde  \lambda^{\dot \alpha}  =0$.  
The latter implies that the spinors satisfy the Weyl equations: 
$0 = p \sigma \tilde \lambda 
\equiv  [p \sigma]_{\alpha \dot \beta} \tilde \lambda^{\dot \beta}$, 
and $0 = p\bar \sigma  \lambda 
\equiv  [p \bar \sigma]^{\dot \alpha \beta} \lambda_{\beta}$. 
The $U(1)$ subgroup of the little group associated with $p^\mu$ acts on the spinors as 
$\lambda\to t^{-1} \lambda$, 
$\tilde \lambda\to t \tilde \lambda$, as this transformation does not change the momentum $p^\mu$. 

On the other hand, a massive momentum, $p^2=m^2$, can be associated with two pairs of two-component spinors 
$\chi_{\alpha}^I$, 
and $\tilde \chi_{\dot \alpha}^I$. 
In addition to the spinor index $\alpha$, the spinors now have the $SU(2)$ little group index  $I =1,2$, so that little group acts as 
$\chi^I \to \chi^L \Omega_L{}^I$, 
$\tilde \chi_I \to \Omega^\dagger_I{}^L \chi_L$.
The little group indices can be raised and lowered in the analogous way as the spinor ones: $\chi_I = \epsilon_{IL} \chi^L$, 
$\tilde \chi^I = \epsilon^{IL} \tilde \chi_L$ (spinor indices are implicit from now on).
The defining relation connecting the massive spinors and momentum are 
$p \sigma =
\chi^I \tilde \chi_I$,
$\chi^{I} \chi_{K}
= \tilde \chi_{K} \tilde \chi^{I}
= m \delta^I_K$. 
It follows that the massive spinors satisfy the Dirac equation, 
$p \bar \sigma \chi^I = m \tilde \chi^I$, 
$p \sigma \tilde \chi^I = m  \chi^I$.  

With the spinor formalism at hand,  scattering amplitudes can be written in the form that makes transparent their little group covariance properties. 
Consider an amplitude 
$\cM[1^h 2^{(J,J_3)}  \dots ]$ containing an incoming massless particle with momentum $p_1^\mu$ and helicity $h$, and an incoming massive particle with momentum $p_2^\mu$, spin $J$, and polarization $J_3$ along some chosen quantization axis. 
The general rules of QFT inform that,  under the massless little group transformation 
$\lambda_1 \to t^{-1} \lambda_1$ and
$\tilde \lambda_1 \to t \tilde  \lambda_1$,
the amplitude must scale as 
$\cM[1^h 2^{(J,J_3)} \dots ] 
\to t^{2h}\cM[1^h  2^{(J,J_3)} \dots ]$.
For massive particles, on the other hand, little group covariance can be realized if the amplitude is constructed out of $2J$ spinors
$\chi_2^{I_i}$ or $\tilde \chi_2^{I_i}$,  such that it is a fully symmetric tensor in the 
$I_1 \dots I_{2J}$ indices. 
Each independent configuration of these indices represents a possible value of polarization: 
$J_3 = J$ corresponds to  
$(I_1,I_2,\dots,I_{2J})=(1,1,\dots,1)$, 
$J_3 = J-1$, corresponds to $(1,1, \dots, 1, 2)$,
$J_3 = J-2$ corresponds to $(1,1, \dots, 1, 2, 2)$, 
and so on. 

We turn to applying this formalism to beta decay. 
We consider the amplitude 
$\cM_{\beta^-} \equiv \cM[1_{\cal N} 2_{\bar {\cal N'}} 3_{e^+} 4_{\nu}]$ for a parent nucleus ${\cal N}$ with spin $J_1$ undergoing $\beta^-$ decay into a daughter nucleus ${\cal N}'$ with spin $J_2$, an electron $e^-$, and an anti-neutrino $\bar \nu$.
For the sake of calculating the amplitude we treat all these particles as incoming, with  momenta denoted by $p_1$, $p_2$, $p_3$, and $p_4$, respectively. 
Amplitudes with outgoing particles can be obtained from $\cM_{\beta^-}$ by crossing symmetry, in particular 
$\cM[1_{\cal N} \to  2_{{\cal N'}} 3_{e} 4_{\bar \nu}] = 
- \cM[1_{\cal N} (-2)_{\bar {\cal N'}} (-3)_{\bar e} (-4)_{\nu}]$.
We will consider $\cM_{\beta^-}$ in the limit of unbroken isospin, in which case 
$p_1^2=p_2^2=m_{\cal N}^2$. 
Moreover, $p_3^2=m_e^2$, while neutrinos are treated as massless, $p_4^2=0$. 
The corresponding spinor variables are  $\chi_1^{I_i}$, $\chi_{2\, O_j}$, $\chi_{3\, K}$, $\lambda_4$,  together with their twiddled counterparts. 
In our conventions, the little group indices of the parent nucleus are up, and those of the daughter nucleus and the electron are down. 
Little group covariance dictates that
$\cM_{\beta^-}$ must be a $2J_1+2J_2+1$ dimensional tensor in the massive little group space, 
$\cM_{\beta^-} \equiv [\cM_{\beta^-}]^{I_1 \dots I_{2J}}_{O_1 \dots O_{2J'} \, K}$, 
which is fully symmetric separately in the 
$I_j$ and $O_j$ indices.
Moreover, it must transform as 
$\cM_{\beta^-} \to t^{-1} \cM_{\beta^-}$ 
under the little group transformations associated with the massless momentum $p_4^\mu$.
 
In the following, we restrict to the case of mixed Fermi--Gamow-Teller allowed transitions where the spins of the parent and daughter nuclei are equal, $J_1 = J_2 \equiv J$. 
In the SM at tree level $\cM_{\beta^-}$ can be factorized into a product of nuclear and $V$-$A$ leptonic currents:  
\begin{align}
\label{eq:BDA_Mbeta-current-SM} 
\cM_{\beta^-} = & 
 - \big [  j_V^\mu  + j_A^\mu  
 \big ] (\tilde \chi_3 \bar   \sigma_\mu  \lambda_4  ) 
,  \end{align}
where $j_{V,A}^\mu$ represent the matrix elements of the vector and axial currents, respectively, between the nuclear states.  
In the limit of unbroken isospin symmetry the terms consistent with parity of the strong interactions contributing up to subleading order in the momentum transfer 
$q^\mu \equiv p_1^\mu + p_2^\mu$ are the following:
\begin{align}
\label{eq:BDA_jVjA-SM}
j_V^\mu = & 
M_F \bigg[
 C_V^i \big [ \chi_2  \sigma^\mu \tilde \chi_1 +  \tilde \chi_2  \bar \sigma^\mu \chi_1 \big ]  - i{ C_M^i -C_V^i \over 2 m_{\cal N} }  
\big [ \chi_2  \sigma^{\mu \nu}  \chi_1 +  \tilde \chi_2  \bar \sigma^{\mu \nu} \tilde  \chi_1 \big ]  q_\nu   
\bigg ] 
\nnl \times &
{ \big [ \chi_2 \chi_1 +  \tilde \chi_2 \tilde \chi_1  \big ]^{2J-1} \over (2 m_{\cal N})^{2 J-1} }   
+ \cO(q^2)
,\nnl 
j_A^\mu = & 
M_F  C_A^i  \big [ \chi_2  \sigma^\mu \tilde \chi_1 -  \tilde \chi_2  \bar \sigma^\mu \chi_1 \big ]  
{ \big [ \chi_2 \chi_1 +  \tilde \chi_2 \tilde \chi_1  \big ]^{2 J-1} \over (2 m_{\cal N})^{2 J-1} }   
+ \cO(q^2) 
. \end{align}
Here, $M_F$ is the Fermi matrix element which, 
for $\beta^\mp$ transitions, is given by 
$M_F = \delta^j_{j'}\delta^{j_3}_{j_3' \pm 1}
\sqrt{j(j+1)-j_3(j_3\pm 1)}$,
where $(j,j_3)$ and $(j',j_3')$ 
are the isospin quantum numbers of the parent and daughter nuclei. 
The Wilson coefficients $C_V^i$, $C_A^i$, and $C_M^i$ are a-priori transition-dependent, which we stress by including the label $i$.  
However, one can show that, in the limit of unbroken isospin, $C_V^i$ is universal for all allowed transitions. 
Furthermore, for the so-called mirror transitions, isospin conservation relates $C_M^i$ to the magnetic momenta 
$\mu_{\cal N}$ and $\mu_{\cal N'}$ of the parent and daughter nuclei: 
${C_M^i \over C_V^i} = 
A {\mu_+ - \mu_-\over \mu_N}$, where $A$ is the mass number for the transition, the nuclear magneton is defined as 
$\mu_N = {e \over 2 m_p}$,
and $\mu_+$ and $\mu_-$ denote the magnetic moments of the nuclei with larger and smaller isospin quantum number $j_3$, respectively.
On the other hand, $C_A^i$ is not fixed by symmetries and currently cannot be determined with enough accuracy from first principles. 
In practice, it is determined phenomenologically from the lifetime and/or correlation measurements. 
We define the antisymmetric combinations of sigma matrices: 
$\sigma^{\mu\nu} = {i \over 2} [ \sigma^\mu \bar \sigma^\nu - \sigma^\nu \bar \sigma^\mu ]$, 
and 
$\bar \sigma^{\mu\nu} = {i \over 2} [ \bar \sigma^\mu  \sigma^\nu - \bar \sigma^\nu  \sigma^\mu ]$. 
To reduce clutter, the little group indices are suppressed in \cref{eq:BDA_jVjA-SM} and in most of the following. 
Implicitly, all $2J$ spinors associated with the parent nucleus carry the upper little group index, $\chi_1^{I_i}$, 
while those associated with the daughter nucleus carry the lower index, $\chi_{2 \, O_i}$. 
Both upper and lower indices are implicitly fully symmetrized, such that the spinorial currents $j_X^\mu \equiv [j_X^\mu]^{I_1 \dots I_{2J}}_{O_1 \dots O_{2J}}$ are  $2J+2J$ dimensional tensors in the little group space, fully symmetric in the upper and lower indices. 
The correct transformation properties of 
$j_V^\mu$ and $j_A^\mu$ under parity follow from the transformations of the spinor bilinear collected in \cref{tab:parity}. 

\begin{table} 
\begin{center}
\begin{tabular}{c|c} 
P{\rm -even}  &   P{\rm -odd} 
\\ \hline 
 $\chi_2 \chi_1 + \tilde \chi_2 \tilde \chi_1$
 &  
$ \chi_2 \chi_1 - \tilde \chi_2 \tilde \chi_1$
\\
$\chi_2  \sigma^0 \tilde  \chi_1 
+ \tilde \chi_2  \bar \sigma^0 \chi_1$
& 
$\chi_2  \sigma^0 \tilde  \chi_1 
-  \tilde \chi_2  \bar \sigma^0  \chi_1 $
\\
$\chi_2  \sigma^k \tilde  \chi_1 
- \tilde \chi_2  \bar \sigma^k \chi_1 $
  &  
$\chi_2  \sigma^k \tilde  \chi_1 + \tilde \chi_2  \bar \sigma^k  \chi_1$
\\
$\chi_2 \sigma^{0k} \chi_1 
-  \tilde \chi_2 \bar \sigma^{0k}   \tilde \chi_1$
 &  
$ \chi_2 \sigma^{0k} \chi_1 
+   \tilde \chi_2\bar \sigma^{0k}   \tilde \chi_1$  
\\
$ \chi_2 \sigma^{jk} \chi_1 
+  \tilde \chi_2\bar \sigma^{jk}   \tilde \chi_1$
 &  
$\chi_2 \sigma^{jk} \chi_1
-   \tilde \chi_2\bar \sigma^{jk}   \tilde \chi_1$
\end{tabular}
 \end{center}
 \caption{
Transformations of spinor bilinears under parity acting as 
$\chi_n^{\alpha \, I} \to
\omega_n \tilde  \chi_{n \, \dot \alpha}^I$, 
where $\omega_1 = 1$, $\omega_2 = -1$. 
 }
  \label{tab:parity}
\end{table}

For $q^2  < 4m_{\cal N}^2$, the massive spinors describing the two nuclei can be related as~\cite{Aoude:2020onz,Cangemi:2023ysz}  
\begin{align}  
\label{eq:BDA_chi2ofchi1}
\chi_2 = &  - {1 \over \sqrt {1 -  {q^2 \over 4m_{\cal N}^2}  }} \bigg  [  \chi_1  -  {q \sigma \over 2 m_{\cal N} } \tilde \chi_1 \bigg ] 
 , \nnl 
\tilde \chi_2 = & {1 \over \sqrt {1 -  {q^2 \over 4m_{\cal N}^2} }}  \bigg [   \tilde  \chi_ 1 -   {q \bar \sigma \over 2 m_{\cal N} }  \chi_1  \bigg ] 
.   \end{align}
It is straightforward  to demonstrate that the above relation leads to the correct Dirac equations for $\chi_2$ and $\tilde \chi_2$. 
\cref{eq:BDA_chi2ofchi1} allows us to simplify the spinor bilinears entering in \cref{eq:BDA_jVjA-SM}. 
Up to quadratic correction in the momentum transfer one finds 
\begin{align}   
\label{eq:BDA_HadronicBilinears}
\chi_2 \chi_1 +  \tilde \chi_2 \tilde \chi_1  = & 
2 m_{\cal N} {\bf 1}  + \cO(q^2)
,\nnl 
\chi_2 \chi_1 - \tilde \chi_2 \tilde \chi_1   = &  -{1\over 2 m_{\cal N}}  \big [  \chi_1 q\sigma    \tilde \chi_1  
+  \tilde \chi_1  q \bar \sigma  \chi_1  \big ]  + \cO(q^2) 
,\nnl
\chi_2 \sigma_\mu  \tilde  \chi_1 +  \tilde \chi_2 \bar \sigma_\mu \chi_1  =  & 
  (p_1 - p_2)_\mu {\bf 1} 
 - {i  \over 2 m_{\cal N}^2 }   
\epsilon_{\mu \nu \alpha \beta} q^\nu p_1^\alpha  \big [ \chi_1\sigma^\beta  \tilde  \chi_1 +   \tilde \chi_1\sigma^\beta   \chi_1 \big ]
+ \cO(q^2) 
,\nnl
\chi_2 \sigma^\mu  \tilde  \chi_1 -  \tilde \chi_2 \bar \sigma^\mu \chi_1  = &  
- \big [ \chi_ 1  \sigma^\mu    \tilde   \chi_1    +   \tilde  \chi_ 1  \bar \sigma^\mu   \chi_1  \big ]  
-    {p_1^\mu  \over  2 m_{\cal N}^2} \big [\chi_1  q  \sigma     \tilde  \chi_1  +   \tilde \chi_1  q \bar \sigma   \chi_1  \big ] 
+ \cO(q^2) 
 ,\nnl
\big[  \chi_2  \sigma_{\mu \nu}  \chi_1  +  \tilde \chi_2  \bar \sigma_{\mu \nu} \tilde  \chi_1  \big ]q^\nu  = & 
  { 1 \over m_{\cal N}  }  
\epsilon_{\mu \nu \alpha \beta} q^\nu p_1^\alpha       \big [ \chi_1\sigma^\beta  \tilde  \chi_1 +   \tilde \chi_1\sigma^\beta   \chi_1 \big ] 
  + \cO(q^2) 
  ,\nnl
\big[  \chi_2  \sigma^{\mu \nu}  \chi_1 -  \tilde \chi_2  \bar \sigma^{\mu \nu} \tilde  \chi_1  \big ]q_\nu  = &  
i {   p_1^\mu - p_2^\mu \over 2 m_{\cal N}  }\big [ \chi_1 q \sigma \tilde  \chi_1 + \tilde  \chi_1q  \bar \sigma \chi_1  \big ]   + \cO(q^2) 
 ,  \end{align}  
where ${\bf 1} \equiv  \delta^I_O$, and where spinors appearing to the left (right) in spinor contractions carry lower (upper) little group indices. 
The second and last bilinears do not appear in \cref{eq:BDA_jVjA-SM}, but we included them for completeness. 
Using these relations, we can rewrite \cref{eq:BDA_jVjA-SM} and the entire amplitude in a more convenient form. 
We organized the amplitude into an expansion in powers of momentum exchange:  
  \begin{align}
\label{eq:BDA_Mbeta}
\cM_{\beta^-} = 
\cM_{\beta^-}^{(0)} +  \cM_{\beta^-}^{(1)} + \cO(q^2)
. \end{align} 
At the leading and subleading order we find 
  \begin{align}
\label{eq:BDA_Mbeta0}
\cM_{\beta^-}^{(0)}  = & 
 - M_F \bigg \{ 
C_V^i (p_1^\mu - p_2^\mu) {\bf 1}^{2J} 
+  {2 m_{\cal N}  \over J}   C_A^i S^\mu  
\bigg \} 
(\tilde \chi_3 \bar   \sigma_\mu  \lambda_4  ) 
, \end{align} 
  \begin{align}
  \label{eq:BDA_Mbeta1}
\cM_{\beta^-}^{(1)}  =  & 
 - { M_F \over m_{\cal N}  J }   \bigg \{  
i C_M^i   
\epsilon^{\mu \nu \alpha \beta} 
q^\nu   p_1^\alpha  S^\beta 
+  C_A^i    (q S)   p_1^\mu 
 \bigg \} (\tilde \chi_3 \bar   \sigma_\mu  \lambda_4  ) 
. \end{align} 
Above, 
${\bf 1}^{2J} \equiv 
\delta^{I_1}_{O_1} \dots \delta^{I_{2J}}_{O_{2J}}$, 
with upper and lower indices fully symmetrized. 
We also introduced the spin vector~\cite{Holstein:2008sx} which, in our conventions, is defined as 
 \begin{align} 
 \label{eq:BDA_spinVector}
[S^\mu]_{O_1 \dots O_{2J} }^{I_1 \dots I_{2J} }  \equiv  &
 - {J \over 2 m_{\cal N} } 
 \big [  \chi_{1\, O_1}  \sigma^\mu    \tilde   \chi_1^{I_1}   +   \tilde  \chi_{1\, O_1}  \bar \sigma^\mu   \chi_1^{I_1} \big ]  
 \delta_{O_2}^{I_2} \dots  \delta_{O_{2J}}^{I_{2J}}
 , \end{align} 
again with upper and lower indices implicitly symmetrized.
From this definition it follows that the spin vector satisfies 
$p_1^\mu S_\mu = 0$, 
and $S^\mu S_\mu = - J(J+1)$.  \\
\cref{eq:BDA_Mbeta0,eq:BDA_Mbeta1} are the starting point to calculate the Coulomb corrections in the following section. 

For $\beta^+$ decay the amplitude can be obtained from the $\beta^-$ one by using CPT and crossing symmetry: 
\begin{align}
\label{eq:BDA_crossing}
\cM[ 1_{\cal N'} 2_{\bar {\cal N}} 3_{e^-} 4_{\bar \nu} ]   =  
    - \cM[(-2)_{\cal N}(-1)_{\bar {\cal N}'} (-3)_{e^+} (-4)_{\nu} ]^* 
.\end{align} 
Applying this to \cref{eq:BDA_jVjA-SM} and expanding in $q$, the analogues of \cref{eq:BDA_Mbeta0,eq:BDA_Mbeta1} for $\beta^+$ decay read 
  \begin{align}
\label{eq:BDA_MbetaPlus0}
\cM_{\beta^+}^{(0)}  = & 
 - M_F \bigg \{ 
 \bar C_V^i (p_1^\mu - p_2^\mu) {\bf 1}^{2J}
+  {2 m_{\cal N}  \over J}   \bar C_A^i S^\mu  
\bigg \} 
(\tilde \chi_3 \bar   \sigma_\mu  \lambda_4  )^* 
, \end{align} 
  \begin{align}
  \label{eq:BDA_MbetaPlus1}
\cM_{\beta^+}^{(1)}  =  & 
 - { M_F \over m_{\cal N}  J }   \bigg \{  
i \bar C_M^i \epsilon^{\mu \nu \alpha \beta} 
q^\nu   p_1^\alpha  S^\beta 
+  \bar C_A^i    (q S)   p_1^\mu 
 \bigg \} (\tilde \chi_3 \bar   \sigma_\mu  \lambda_4)^*  
. \end{align} 
We can see that $\cM_{\beta^+}$ in the SM limit is obtained from $\cM_{\beta^-}$ by conjugating all leptonic currents and Wilson coefficients.\footnote{%
For non-SM scalar and tensor current introduced later in \cref{sec:NONSM} the procedure implies also a sign flip of the corresponding Wilson coefficients $C_{S,T}^i$.
}
In the same way, \cref{eq:BDA_crossing} allows one to quickly derive the form of Coulomb corrections for $\beta^+$ transitions starting from the $\beta^-$ results in the next section.

\section{Coulomb corrections from unitarity} 
\label{sec:CC}

In this section we discuss Coulomb corrections to the $\beta$ decay amplitude in \cref{eq:BDA_Mbeta}. 
At one loop, this amplitude receives contributions due to photon exchange between charged external particles. 
The Coulomb corrections are defined as the part of these contributions originating from the intermediate particles in the loop going on shell. 
For real Wilson coefficients $C_X^i$ in \cref{eq:BDA_Mbeta} this corresponds to the imaginary part of the loops; 
more generally, Coulomb corrections are related to the discontinuity of the one-loop amplitude when some kinematic variables crosses the threshold for pair production of intermediate particles.   
Via unitarity, such discontinuities can be expressed as products of tree-level amplitudes integrated over the phase space of intermediate states. 
Therefore, unitarity allows one to calculate the Coulomb corrections  without  performing any actual loop integrals, and this is the way we will proceed in the following. 
Within the physical kinematics of beta decay one can verify that, at one loop,  the only discontinuity appears on the positive $u$ axis for $u>u_0$, with $u \equiv (p_1+p_4)^2$ and  the threshold at 
$u_0 = (m_{\cal N} + m_e)^2$. 
This corresponds to the final-state interaction contribution to the one loop amplitude where a photon is exchanged between the final-state electron and daughter nucleus. 
Unitarity dictates that the discontinuity can be expressed as 
\begin{equation}
\label{eq:CC_Master_Disc}
\text{Disc}_2^u {\cal M} \left[ 1_{\cal N} 2_{\bar\N'} 3_{e^+} 4_\nu \right] =  
i \int  \text{d}\Pi_{XY} {\cal M}_{\beta^-}  \left[ 1_{\cal N} (-Y)_{\cal \bar{N'}} (-X)_{e^+} 4_\nu\right]   
{\cal M}_{\text{em}} \left[ Y_{\cal N'} X_{e^-} 2_{\bar{\cal N'}} 3_{e^+} \right].
\end{equation}
Here, 
$\text{d}\Pi_{XY}$ denotes the two-body phase space of the intermediate electron and nucleus states labeled by $X$ and $Y$. 
Implicitly, these particles carry little group indices (lower in ${\cal M}_{\beta^-}$,  upper in ${\cal M}_{\text{em}}$), which are contracted between each pair and summed over. 
A convenient parametrization of the phase space, which reduces it to trivial polynomial integration,  is presented in~\cref{app:INT}. Next, ${\cal M}_{\beta^-}$ is the tree-level beta decay amplitude in~\cref{eq:BDA_Mbeta-current-SM}. 
Finally, 
${\cal M}_{\text{em}}$ is the tree-level amplitude describing electromagnetic re-scattering of the daughter nucleus and electron.  
The Coulomb corrections to the beta decay amplitude are then given by one half of the discontinuity in~\cref{eq:CC_Master_Disc}. 

It is convenient to organize the discontinuity as follows:  
\begin{align}
\label{eq:CC_Master_Discij}
\text{Disc}_2^u {\cal M} \left[ 1_{\cal N} 2_{\bar\N'} 3_{e^+} 4_\nu \right] 
= & 
\sum_{i,j}  D^{(i,j)}
, \nnl 
D^{(i,j)} \equiv & 
i \int  \text{d}\Pi_{XY} {\cal M}_{\beta^-}^{(i)}  \left[ 1_{\cal N} (-Y)_{\cal \bar{N'}} (-X)_{e^+} 4_\nu\right]   
{\cal M}_{\text{em}}^{(j)} \left[ Y_{\cal N'} X_{e^-} 2_{\bar{\cal N'}} 3_{e^+} \right]
. \end{align}
Above, ${\cal M}_{\beta^-}^{(i)}$ contains $\cO(q^i)$ terms in the expansion of the beta decay amplitude in powers of momentum exchange,  
cf.~\cref{eq:BDA_Mbeta}. 
${\cal M}_{\text{em}}^{(j)}$ refers to the analogous expansion of the electromagnetic scattering amplitude. 
Explicitly, at the leading order we have 
\begin{align}
 \label{eq:CC_Mem0}
&\cM_{\rm em}^{(0)} \big [1_{\cal N} 2_{e^-}  3_{\bar{\cal N}}   4_{e^+}  \big ]    = 
-  {2  q_e Z e^2 \over (p_1 + p_3)^2 }
  \big [ \chi_4 p_3 \sigma \tilde \chi_2 + \tilde \chi_4 p_3 \bar \sigma  \chi_2 \big ]  {\bf 1}^{2J}   
, \end{align}
where $e \simeq 0.3$ is the electromagnetic coupling constant, 
$Z$ is the nuclear charge,  
and $q_e = -1$ is the electron charge. 
At the next-to-leading order,
 \begin{align}
 \label{eq:CC_Mem1}
\cM_{\rm em}^{(1)} \big [1_{\cal N} 2_{e^-}  3_{\bar{\cal N}}   4_{e^+}  \big ]    = &  
{2  i q_e e \mu_{\cal N}  \over  J (p_1 + p_3)^2 }  
 \epsilon_{  \mu \nu \rho  \alpha}  ( p_1^\nu + p_3^\nu ) p_1^\rho  S^\alpha
  \big [  \chi_4 \sigma^\mu \tilde  \chi_2  +  \tilde \chi_4 \bar \sigma^\mu  \chi_2     \big ]  
  ,   \end{align}
where  $\mu_{\cal N}$ is the nuclear magnetic moment with dimension 
$[\mu_{\cal N}] = {\rm mass}^{-1}$.
We are interested in the discontinuity up to linear order in $q$. 
Therefore we need to calculate 
$D^{(i,j)}$ with $i+j \leq 1$.  

We start from the leading order  contributions to the discontinuity.
$D^{0,0}$ is obtained by plugging \cref{eq:BDA_Mbeta0} and \cref{eq:CC_Mem0} into \cref{eq:CC_Master_Discij}.
For the presentation purpose we split 
$D^{0,0} = 
C_V^i D^{0,0}_V + C_A^i D^{0,0}_A$. 
Simplifying the contractions of spinors describing the intermediate particles,   performing the phase space integral, and expanding the result in powers\footnote{%
In this expansion we count 
$p_1 \sim p_2 \sim \cO(m_{\cal N})$, and the remaining momenta as $\cO(m_{\cal N}^0)$. } 
of $1/m_{\cal N}$ 
we find
\begin{align} 
\label{eq:CC_D00}
D_V^{0,0}= &  
  - {i  M_F q_e Z' e^2  
 \over  8 \pi m_\N \mathcal{K}_e  } 
{\bf 1}^{2J} \bigg \{ 
(p_1^\mu - p_2^\mu)
(\tilde\chi_3\bar\sigma_\mu  \lambda_4)
 m_e^2  
- 4   m_e   (  \chi_3 \lambda_4 ) \big ( m_{\cal N}^2 -  p_1 p_3  \big ) 
\bigg \}  
\nnl & 
-   i  \phi   
 M_F {\bf 1}^{2J}
(p_1^\mu - p_2^\mu)
(\tilde\chi_3\bar\sigma_\mu  \lambda_4)
+ \cO(m_{\cal N}^{-1})
, \nnl 
D_A^{0,0}= &    
    {i M_F q_e Z' e^2  
\over  2  \pi  J m_\N \mathcal{K}_e   } 
 S^\mu \bigg \{ 
 m_{\cal N}    (\tilde \chi_3 \bar   \sigma_\mu  \lambda_4  )      
 m_e^2 
-    {m_e \over  m_{\cal N}}  (\chi_3 p_2 \sigma \bar \sigma_\mu  \lambda_4 )  
\big (  m_{\cal N}^2  + p_1 p_3  \big )  
\bigg \} 
\nnl & 
-    i \phi {2  m_{\cal N} M_F \over J} 
 S^\mu  (\tilde \chi_3 \bar   \sigma_\mu  \lambda_4  )   
+ \cO(m_{\cal N}^{-1})
,  \end{align}
where $Z'$ is the charge of the daughter nucleus, 
$\mathcal{K}_e=\sqrt{\frac{(p_1 p_3)^2}{m_\N^2}-m_e^2}$ is the magnitude of the electron 3-momentum in the rest frame of the parent nucleus, 
and $\phi \equiv { (-q_e) Z' e^2 
 \over  4 \pi m_\N \mathcal{K}_e  }  
   (p_2 p_3) \big [ 
  \int { d \alpha   \over  \alpha}  -1   
 \big ] $. 
Here, $\int { d \alpha   \over  \alpha} \equiv 
\int_0^{\alpha_0} { d \alpha   \over  \alpha}$  denote an IR divergent integral over the phase space parameter $\alpha$, 
with the upper integration limit $\alpha_0$ defined in \cref{app:INT}. 
The divergence is controlled by the Weinberg's soft theorem, and its contribution to the beta decay amplitude can be represented as an universal shift of all Wilson coefficients in \cref{eq:BDA_Mbeta0,eq:BDA_Mbeta1}:
$\delta C_X^i =   i \phi C_X^i$. 
For the lifetime and correlation coefficients, 
which always depend on the Wilson coefficients via the combinations $C_X^i \bar C_{X'}^i$, the effect of such a universal imaginary shift is null at $\cO(e^2)$ at which we perform out calculation.  
Therefore, the $\int {d\alpha\over \alpha}$ terms in~\cref{eq:CC_D00} can be ignored for all practical purpose. 
The $\cO(m_{\cal N}^1)$ terms in \cref{eq:CC_D00} encode complete information about leading order Coulomb corrections to all correlation coefficients, which were originally listed in Ref.~\cite{Jackson:1957auh}, and which contribute to $D$ only beyond the SM. 
On the other hand, the $\cO(m_{\cal N}^0)$ terms in \cref{eq:CC_D00} will contribute to $D$ in the SM limit.  

We move to discuss the discontinuity at the subleading order in $q$, which is contained in $D^{1,0}$ and $D^{0,1}$. 
The former is obtained by  plugging \cref{eq:CC_Mem0,eq:BDA_Mbeta1} into \cref{eq:CC_Master_Discij}. 
Once again we split 
$D^{1,0} = C_M^i D^{1,0}_M + C_A^i D^{1,0}_A$.  
We get 
   \begin{align}
D^{1,0}_M =  &   
 {  M_F  q_e Z' e^2\over 8  \pi  \mathcal{K}_e    m_{\cal N}^2   J } 
\bigg \{ 
\epsilon_{\mu \nu \alpha \beta }   p_1^\mu S^\nu  \bigg  [   (\tilde \chi_3  \bar \sigma^\alpha  \lambda_4 )   \bigg (    2 (p_1 p_3) p_3^\beta   
-  \mathcal{K}_e^2  \big [  c_3 p_3^\beta + c_4 p_4^\beta \big ]  \big[  c_3 (p_1p_3)   - c_4 (p_1 p_4)\big ]    \bigg )
   \nnl  \hspace{-2.6cm}   &  
 +  m_e  (\chi_3 p_2 \sigma  \bar \sigma^\alpha  \lambda_4 )  \big [    
  -2  p_4^\beta  
     + \mathcal{K}_e^2  c_3  \big ( c_3 p_3^\beta + c_4 p_4^\beta \big )  
  \big ] 
+ 2 (\tilde \chi_3  p_1 \bar \sigma   \lambda_4 ) p_3^\alpha  p_4^\beta   
 \mathcal{K}_e^2  c_3  c_4 
\bigg ] 
    \nnl    & 
-  i (S p_3 )    \mathcal{K}_e^2  m_{\cal N}^2   c_\epsilon^2  \bigg (
   (\tilde \chi_3  p_1 \sigma   \lambda_4 )     (p_1 p_4 )  (p_3 p_4) 
+    m_e ( \chi_3  \lambda_4 )    (p_1 p_4)^2    
\bigg )
    \nnl   & 
-  i (S p_4) 
  \mathcal{K}_e^2   m_{\cal N}^2   c_\epsilon^2   \bigg (
   (\tilde \chi_3  p_1 \bar \sigma   \lambda_4 ) \big[      (p_1  p_3) (p_3 p_4)  -  m_e^2  (p_1 p_4)    \big] 
\notag \\
&\! + \! m_e ( \chi_3  \lambda_4 )  \big [m_{\cal N}^2   (p_3 p_4)   - (p_1 p_3)  (p_1 p_4)   \big ]     
\bigg )
\bigg \} 
\nnl & 
- i \phi {  i M_F  \over    m_{\cal N}   J } 
\epsilon_{\mu \nu \alpha \beta }   
p_1^\mu S^\nu 
   (\tilde \chi_3  \bar \sigma^\alpha  \lambda_4 )  q^\beta   
   + \cO(m_{\cal N}^{-1}) 
   \nonumber 
 ,  \end{align} 
\begin{align}
\label{eq:CC_D10}
D^{1,0}_A =  &          
{ M_F q_e Z e^2   
\over 8 \pi   \K_e  J }   \bigg \{   
- i (S_1 p_3)    \bigg [   
{ ( \tilde \chi_3 p_1 \bar \sigma    \lambda_4  )  \over m_{\cal N}^2}     \big [
      2 (p_1 p_3)        
      -  c_3  \big (  c_3 (p_1 p_3) + c_4 (p_1 p_4) \big )  \mathcal{K}_e^2  
     \big  ]                   
      -   m_e ( \chi_3  \lambda_4  )     c_3^2    \mathcal{K}_e^2     \bigg ] 
                  \nnl  \hspace{-2.5cm}   &  
 +  i (S_1 p_4)  \bigg [  
 {   ( \tilde \chi_3 p_1 \bar \sigma    \lambda_4  )  
 \over m_{\cal N}^2} c_4  \big (  c_3 (p_1 p_3) + c_4 (p_1 p_4) \big )  \mathcal{K}_e^2   
-   m_e ( \chi_3  \lambda_4  )       \big (    2   -  c_3 c_4  \mathcal{K}_e^2  \big )        
  \bigg ] 
\nnl  \hspace{-2.5cm}   &  
 -    c_\epsilon^2   \mathcal{K}_e^2    \epsilon_{\mu \nu \alpha \beta}  p_1^\mu S_1^\nu p_3^\alpha p_4^\beta 
   \bigg [ 
  ( \tilde \chi_3  p_1 \bar \sigma  \lambda_4  )   (p_3 p_4)   
   +   m_e   (\chi_3  \lambda_4  )   (p_1 p_4 )  
   \bigg ] 
        \bigg \}  
 \nnl & 
 -  i \phi { M_F   
\over   J m_{\cal N}  }  
 (S_1 q) ( \tilde \chi_3 p_1 \bar \sigma    \lambda_4  )  
 + \cO(m_{\cal N}^{-1}) 
 ,  \end{align}   
where the coefficients $c_{3,4,\epsilon}$ are related to our parametrization of internal momenta in the phase space integral, and are defined in \cref{app:INT}.

Finally we deal with $D^{0,1}$, which is obtained by plugging \cref{eq:BDA_Mbeta0,eq:CC_Mem1} into \cref{eq:CC_Master_Discij}.
Splitting 
$D^{0,1} = 
C_V^i D^{0,1}_V +  C_A^i D^{0,1}_A $, we find
\begin{align}
 D^{0,1}_V = & \; 
    i \frac{(-q_e) e \mu_{\cal N'} M_F}{4 \pi J m_\N \K_e} 
    \bigg\{ 
        2 i \epsilon_{\nu \alpha \mu \rho} p_1^\nu  S^\alpha p_3^\rho \bigg[
            -(p_1 p_3) (\tilde \chi_3 \bar \sigma^\mu \lambda_4) 
            + m_e (\chi_3 p_2 \sigma \bar \sigma^\mu \lambda_4)
        \bigg] \nnl
    &  + 2 i \mathcal{K}_e^2 \epsilon_{\nu \alpha \mu \rho} p_1^\nu  S^\alpha p_3^\mu p_4^\rho 
        (\tilde \chi_3 p_1 \bar \sigma \lambda_4) c_3 c_4 \nnl
    &  + i \mathcal{K}_e^2 \epsilon_{\nu \alpha \mu \rho} p_1^\nu S^\alpha 
         \big[c_3  p_3^\rho + c_4  p_4^\rho \big] \bigg[ 
            \big[ c_4 (p_1 p_4) - c_3 (p_1 p_3) \big] 
            (\tilde \chi_3 \bar \sigma^\mu \lambda_4) \nnl
    & \quad + c_3  m_e (\chi_3 p_2 \sigma \bar \sigma^\mu \lambda_4)
        \bigg]  \nnl
    & +  m_{\cal N}^2 \mathcal{K}_e^2 c_\epsilon^2 (S p_3) (p_1 p_4) 
        \bigg[
            (p_3 p_4) (\tilde \chi_3 p_1 \bar \sigma \lambda_4) 
            + m_e (p_1 p_4) (\chi_3 \lambda_4)
        \bigg] \nnl
    & +  m_{\cal N}^2 \mathcal{K}_e^2 c_\epsilon^2 (S p_4) 
        \bigg[
            \big[(p_1 p_3)(p_3 p_4) - m_e^2 (p_1 p_4)\big] 
            (\tilde \chi_3 p_1 \bar \sigma \lambda_4) \nnl
    & \quad + m_e \big[m_{\cal N}^2 (p_3 p_4) - (p_1 p_3)(p_1 p_4)\big] 
            (\chi_3 \lambda_4)
        \bigg]    \bigg\}
        + \cO(m_{\cal N}^{-1})
\nonumber
,\end{align}
\begin{align}
\label{eq:CC_D01} 
D^{0,1}_A = & \; 
    -\frac{q_e e \mu_{\cal N'} M_F}{\pi J^2} 
    \epsilon_{\mu \nu \rho \alpha} p_1^\nu \bar S^\alpha S_M 
    \bigg\{
    \frac{ p_3^\rho}{2 \K_e} 
\bigg [  
 \frac{(p_1 p_3)}{m_{\cal N}^2} 
(\tilde \chi_3 \bar \sigma^\mu p_2 \sigma \bar \sigma^M \lambda_4)
+ m_e (\chi_3 \sigma^\mu \bar \sigma^M \lambda_4) 
\bigg ] 
\nnl
    & \quad + \frac{\mathcal{K}_e (c_3 p_3 + c_4 p_4)^\rho}{4} 
        \bigg[
           \frac{c_3 (p_1 p_3) + c_4 (p_1 p_4)}{m_\N^2} (\tilde \chi_3 \bar \sigma^\mu p_2 \sigma \bar \sigma^M \lambda_4) 
\nnl
    & \qquad 
+ 2 c_3 p_3^\mu (\tilde \chi_3 \bar \sigma^M \lambda_4)
+ 2 c_4 p_4^M (\tilde \chi_3 \bar \sigma^\mu \lambda_4)
+ c_3 m_e (\chi_3 \sigma^\mu \bar \sigma^M \lambda_4) 
        \bigg] \nnl
    & \quad 
+ i \frac{\mathcal{K}_e c_\epsilon^2 \epsilon^{\rho PRL} p_1^P p_3^R p_4^L}{4} 
\bigg[ (p_3 p_4) (\tilde \chi_3 \bar  \sigma^\mu   \sigma^M  p_2 \bar\sigma \lambda_4) \nnl
& \qquad - (p_1 p_4) \big[
2 p_3^\mu (\tilde \chi_3 \sigma^M \lambda_4) - 2 p_3^M (\tilde \chi_3 \bar \sigma^\mu \lambda_4) + m_e (\chi_3 \bar \sigma^\mu \bar \sigma^M \lambda_4)
            \big]  \nnl
    & \qquad - 2 i \epsilon^{MBCD} p_1^B p_3^C p_4^D (\tilde \chi_3 \bar \sigma^\mu \lambda_4)
        \bigg]
    \bigg\}
     + \cO(m_{\cal N}^{-1}) 
. \end{align}
For the sake of power counting note that $\mu_{\cal N'} \sim \cO(m_{\cal N}^{-1})$.
The product of two spin vectors in $D^{0,1}_A$ can be transformed using \cref{eq:spin_multiplication}, but that would lead to a longer expression.


\section{Application to $T$-odd correlations} 
\label{sec:D}

We can now use the information gathered in the previous section to compute Coulomb corrections to the $D$ parameter, as well as other $T$-odd correlation coefficients. 
The observables we are interested in are customarily defined via the differential distribution of the decay width in the rest frame of the parent nucleus, after summing over polarizations of the daughter nucleus and electron: 
    \begin{align} 
\label{eq:D_correlationFromGamma}
{d \Gamma \over   d E_e d  \Omega_e d \Omega_\nu } \supset & 
\xi_0 { M_F^2 k_e E_\nu \over 64 \pi^5 J }   
(\boldsymbol{k_e} \times \boldsymbol{k_\nu}) \cdot \bigg \{ 
\boldsymbol{J} {\color{red} D(E_e) }  \notag \\
& -  \boldsymbol{j}  { J (J+1) - 3 (\boldsymbol{J} \boldsymbol{j} )^2 
 \over  J+1 }   
   \bigg [ 
   {\color{red} \tilde  c_3(E_e) } {(\boldsymbol{j}  \boldsymbol{k_e})   \over   E_e } 
   +    {\color{red} \tilde c_4(E_e) } { (\boldsymbol{j}  \boldsymbol{k_\nu})     \over  E_\nu}  
   \bigg ] 
\bigg \} 
. \end{align}       
Connecting to the previously used relativistic notations, 
$p_1 = (m_{\cal N},\boldsymbol{0})$, 
$p_3 = -(E_e, \boldsymbol{k_e})$, 
$p_4 = -(E_\nu, \boldsymbol{k_\nu})$, 
and $k_e = |\boldsymbol{k_e}|$. 
Furthermore, $\boldsymbol{J}$ is the polarization of the parent nucleus, $\boldsymbol{j}$ is the unit vector in the direction of polarization, 
and the normalization factor is 
$\xi_0 = |C_V^i|^2 
+  {J+1 \over J} |C_A^i|^2$ in the SM. 
The $D$ parameter was defined by Jackson, Treiman and Wyld in Ref.~\cite{Jackson:1957zz}. 
The other two $T$-odd coefficients~\cite{Holstein:1974zf,Falkowski:2021vdg}, which we call $\tilde c_3$ and $\tilde c_4$, arise at the next-to-leading order in momentum exchange for transitions with $J\geq 1$.\footnote{%
Compared to the notation used in in Ref.~\cite{Falkowski:2021vdg}, 
$\tilde c_3 =   - { E_e \over E_\nu}c_4$, 
 $\tilde c_4 =  - c_3$ in parent's rest frame. }

In practice, taking advantage of our fully relativistic formalism, it is more convenient to define the correlation coefficients at the level of the amplitude squared:
        \begin{align}  
\label{eq:D_correlationFromMsq}
        \sum |\cM_{\beta^\mp}|^2  \supset &   
 - { 8  M_F^2  \xi_0  m_{\cal N}   \over J} 
   \epsilon_{\mu \nu \alpha  \beta } 
   p_1^\mu p_3^\alpha  p_4^\beta 
\bigg \{    
   S^\nu D
- S^{\nu}_{\;\rho} 
{2 m_{\cal N }  \over J+1 } \bigg[ 
\tilde c_3   
\frac{p_3^\rho}{(p_1 p_3)} 
+ \tilde c_4 
\frac{p_4^\rho}{(p_1 p_4)} \bigg]
     \bigg \} 
,  \end{align}  
where the sum goes over polarization of the daughter nucleus and electron, 
and the spin tensor of the parent nucleus is defined in \cref{app:CON}. 
The correlations $D$, $\tilde c_3$, $\tilde c_4$ are functions of Lorentz-invariant momentum contractions $(p_1 p_3)$, $(p_1 p_4)$, and $(p_3 p_4)$.  
In the rest frame of the parent nucleus and at the order we are working they are equivalent to the respective correlation $D(E_e)$, $\tilde c_3(E_e)$, $\tilde c_4(E_e)$ defined in the standard way via~\cref{eq:D_correlationFromGamma}.\footnote{%
More generally, the two sets of correlation coefficients differ by the effects due to subleading corrections to the 3-body decay phase space. 
Furthermore, if the Lorentz-invariant coefficients depend on $(p_3 p_4)$, the corresponding rest frame coefficients pick up a non-trivial angular dependence, 
e.g. $D(E_e) \to D(E_e) 
+{\boldsymbol{k_e}\boldsymbol{k_\nu} \over E_e E_\nu} D'(E_e)$. 
Neither of these effects is relevant for the following discussion of Coulomb corrections.
} 

We compute the amplitude squared taking into account the tree-level contributions and Coulomb corrections contributing up to subleading order in momentum exchange: 
\begin{equation}
\label{eq:D_Msq}
|\cM_{\beta^\mp} |^2  \subset 
|\cM_{\beta^\mp}^{(0)}|^2  
+ \bigg \{ 
\cM_{\beta^\mp}^{(0)} \bar\cM_{\beta^\mp}^{(1)}  
+ 
\cM_{\beta^\mp}^{(0)}  
\sum_{i+j \leq 1} \bar\cM_{\beta^\mp}^{C(i,j)}
+\hc \bigg \}  
, \end{equation}  
where $\cM_{\beta^\mp}^{(n)}$ are defined in \cref{eq:BDA_Mbeta0,eq:BDA_Mbeta1}, 
and the Coulomb corrections are $\cM_{\beta^\mp}^{C(i,j)} = {1\over 2} D^{(i,j)}$ in terms of the discontinuities calculated in \cref{eq:CC_D00,eq:CC_D10,eq:CC_D01}. 
Given these amplitudes we perform the polarization sum using the sum rules in \cref{app:CON}.\footnote{The analysis can be easily generalized to include the possibility of  polarized electrons 
upon proper modification of the sum rules.}
We can then extract the coefficients of interest by identifying the corresponding Lorentz-invariant kinematic structure in the summed amplitude squared. 

The first two terms in \cref{eq:D_Msq} give the tree-level contributions to the $D$ parameter up to subleading order in $1/m_{\cal N}$ expansion~\cite{Jackson:1957zz,Falkowski:2021vdg}: 
\begin{align}
D^{\rm tree} \xi_0  = &
\im \bigg \{ 
- 2 C_V^i \bar C_A^i
\mp {(p_1 p_3) - (p_1 p_4) 
\over m_{\cal N}^2  }
C_V^i \bar C_M^i 
+ {(p_1 p_3) + (p_1 p_4) 
\over 2 J m_{\cal N}^2 }
C_A^i   \bar C_M^i \bigg \} 
+ \cO(m_{\cal N}^{-2})
, \nnl
\tilde c_3^{\rm tree}  \xi_0  = &  
\pm {(p_1 p_3)  \over 2  m_{\cal N}^2}  
{J+1 \over J}   \im  [C_A^i  \bar C_M^i] 
+ \cO(m_{\cal N}^{-2}) 
, \nnl
\tilde c_4^{\rm tree}  \xi_0  = &  
\mp { (p_1 p_4)  \over 2 m_{\cal N}}  
{J+1 \over J}  \im  [C_A^i  \bar C_M^i] 
+ \cO(m_{\cal N}^{-2}) 
, \end{align}
where, from here onwards, the upper (lower) signs refer to $\beta^-$ ($\beta^+$) decay.
In all these cases, the $T$-odd  correlations are zero for real Wilson coefficients.  
CP-violation in the SM eventually does contribute to imaginary parts of $C_V^i$ and $C_A^i$ at higher loop levels, but that is very suppressed and impossible to observe in a foreseeable future~\cite{Herczeg:1997se}. 
At the same time, CP-violating contributions from beyond the SM are very constrained by a host of precision measurements, from EDMs to LHC collider observables~\cite{Falkowski:2022ynb}.

The main phenomenological importance of Coulomb corrections lies in the fact that they contribute to the $T$-odd correlations even in the absence of CP violation, that is  for real Wilson coefficients in \cref{eq:BDA_Mbeta0,eq:BDA_Mbeta1}. 
Specifically, we find that the Coulomb corrections contribute to the $D$ parameter as 
\begin{align}
 D^{\rm C} \xi_0 =&
 - {e^2 q_e Z'  
\over 16 \pi  m_{\cal N} {\cal K}_e } 
\re \bigg \{ 
(C_A^i  \pm  C_M^i) \bigg [ 
 \bar C_V^i \bigg ( 
 3 m_e^2  + { (p_1 p_3)^2 \over m_{\cal N}^2}  \bigg ) 
 \mp {\bar C_A^i\over 2 J} \bigg ( 
  m_e^2   + 3 { (p_1 p_3)^2 \over m_{\cal N}^2}  \bigg) 
\bigg ] 
\bigg \}
\nnl &
+ {3 e q_e  \mu_{\cal N'}  
\over 8 \pi J  }  {\cal K}_e 
\re \bigg \{ 
\pm J |C_V^i|^2 
\mp (J+1) |C_A^i|^2
-C_V^i \bar C_A^i \bigg \} 
\end{align}
In the rest frame, this agrees with the results in the previous literature calculated for $J=1/2$ in  Ref.~\cite{Callan:1967zz} and generalized to arbitrary spin in Ref.~\cite{Holstein:1974zf}.\footnote{%
For neutron decay, additional subleading isospin-breaking corrections were calculated in Ref.~\cite{Cirigliano:2022hob}. 
These are included in the numerical evaluation in~\cref{tab:results} as a shift of the $C_M^i$ Wilson coefficient, which results in a $\cO(10^{-7})$ shift of the $D$ parameter.}

Similarly, we can compute Coulomb corrections to the $T$-odd  $\tilde c_3$ and $\tilde c_4$ correlation coefficients. 
We find 
\begin{align}
\tilde c_3^{\rm C} \xi_0  =& 
{3 e^2 Z' q_e \over 32 \pi } 
{J+1 \over J}
{(p_1 p_3)^2 \over m_{\cal N}^3
 {\cal K}_e }
\bigg \{ 
|C_A^i|^2 \pm  \re [C_A^i \bar C_M^i]
\bigg \}
+ \cO(m_{\cal N}^{-2})  
,\nnl 
\tilde c_4^{\rm C} \xi_0 = &  \cO(m_{\cal N}^{-2}) 
.\end{align}

In \cref{tab:results} we give the numerical values of the $T$-odd correlation coefficients predicted in the SM limit for a number of mirror transitions. 
We assume that the vector Wilson coefficients  are universal for all transitions, $C_V^i \to C_V$, as in the case in the limit of unbroken isospin symmetry.
The $D$ and $\tilde c_3$ parameters are strongly transition-dependent, but their magnitude is generally of order $10^{-4}$.
On the other hand, $\tilde c_4$ is not generated in the SM limit at the order we perform our calculation.

\begin{table}[]
    \centering
    \begin{tabular}{c|c|c|c|c|c|c|c|}
Parent  & 
$J$ &  $C_A^i [v^{-2}]$ & $C_M^i [v^{-2}]$ 
& $D [\times 10^{-4}]$ 
& $\tilde c_3 [\times 10^{-4}]$   
 \\ \hline
n   
& 1/2  &-1.2575 & 4.6085 & 
$0.11 {k_e \over k_e^{\rm max}} 
+ 0.02 {k_e^{\rm max} \over k_e}$ 
&   x  
\\ \hline
${}^{17}{\rm F}$ &
5/2 & -1.0786 & 110.9 
& $-0.19 {k_e \over k_e^{\rm max}} 
- 0.15 {k_e^{\rm max} \over k_e}$ 
&  $-2.02 {E_e \over k_e}{E_e \over E_e^{\rm max}}$ 
\\ \hline
${}^{19}{\rm Ne}$ &
1/2 & 0.9115 & -84.54 & 
$2.31 {k_e \over k_e^{\rm max}} 
+ 0.17 {k_e^{\rm max} \over k_e}$ 
&   x 
\\ \hline
${}^{23}{\rm Mg}$  
&  3/2 & 0.4183 & -62.44 &
$2.64 {k_e \over k_e^{\rm max}} 
+ 0.16 {k_e^{\rm max} \over k_e}$ 
& $-1.73 {E_e \over k_e}{E_e \over E_e^{\rm max}}$  
\\\hline 
${}^{35}{\rm Ar}$ 
& 3/2 & -0.2164 & -6.545 
& $0.43 {k_e \over k_e^{\rm max}} 
+ 0.01 {k_e^{\rm max} \over k_e}$ 
& $0.17 {E_e \over k_e}{E_e \over E_e^{\rm max}}$ 
\\\hline 
${}^{37}{\rm K}$  
&  3/2 & 0.4419 & -34.39 
&  $2.32 {k_e \over k_e^{\rm max}} 
+ 0.05 {k_e^{\rm max} \over k_e}$ 
& $-1.59 {E_e \over k_e}{E_e \over E_e^{\rm max}}$ 
\\\hline 
${}^{39}{\rm Ca}$   
&  3/2 & -0.5042 & 24.22 
& $-0.47 {k_e \over k_e^{\rm max}} 
- 0.02 {k_e^{\rm max} \over k_e}$
& $-1.28 {E_e \over k_e}{E_e \over E_e^{\rm max}}$ 
\\ \hline
    \end{tabular}
    \caption{
Evaluation of the $T$-odd correlation coefficients generated by Coulomb corrections in the SM in selected mirror transitions. 
The central values of the Wilson coefficients $C_A^i$ and  $C_V^i = 0.9857 v^{-2}$ with $v\equiv (\sqrt 2 G_F)^{-1/2} \simeq 246.22$~GeV are obtained from a fit to beta decay data along the lines of Ref.~\cite{Falkowski:2020pma}.
The Wilson coefficients $C_M^i$ is calculated using isospin symmetry from the input of magnetic moment of the parent and daughter nuclei. 
    }
    \label{tab:results}
\end{table}

\section{Towards NLO Coulomb corrections beyond SM} 
\label{sec:NONSM}

In this section we generalize the discussion of the previous sections, allowing for non-standard contributions to the beta decay amplitude.
Namely, we supplement the tree-level leading-order $\beta$ decay amplitude with scalar and tensor interactions.  
These arise from integrating out certain kinds of heavy non-SM particle (e.g. leptoquarks) coupled to the first generation leptons and quarks.
Accordingly, they could also descend from certain dimension-6 operators in the SMEFT or WEFT (the effective theory below the electroweak scale). 
In the Lorentz-invariant and little-group covariant language they can be 
parametrized as 
  \begin{align} 
\Delta_{\rm BSM} \cM_{\beta^-} = & 
 - M_F \bigg \{ 
C_S^i \big [ \chi_2 \chi_1 +  \tilde \chi_2 \tilde \chi_1  \big ] (\chi_3  \lambda_4 )   
\nnl & 
+  {1 \over 2 } C_T^i 
\big [  \chi_2  \sigma^{\mu \nu}  \chi_1 -  \tilde \chi_2  \bar \sigma^{\mu \nu} \tilde  \chi_1 \big ]  
  ( \chi_3 \sigma_{\mu \nu} \lambda_4 )   
 \bigg \}  { \big [ \chi_2 \chi_1 +  \tilde \chi_2 \tilde \chi_1  \big ]^{2J-1} \over (2 m_{\cal N})^{2J-1} } 
  . \end{align}    
Expanding the nuclear spinor bilinear using \cref{eq:BDA_HadronicBilinears} one finds at the leading and subleading orders in momentum exchange
\begin{align}
\label{eq:beta_bsm}
\Delta_{\rm BSM} \cM_{\beta^-}^{(0)}  = & 
 - 2 M_F \bigg \{ 
  m_{\cal N}  C_S^i  {\bf 1}^{2J}  (  \chi_3 \lambda_4 )  
 + { S^\mu  \over J}   C_T^i  (\chi_3 p_1 \sigma  \bar \sigma_\mu  \lambda_4 ) 
\bigg \} 
, \end{align}
\begin{align}
\label{eq:beta_bsm1}
\Delta_{\rm BSM} \cM_{\beta^-}^{(1)}  = & 
-   {M_F \over J}   C_T^i  \bigg \{ 
  { 1 \over 2}  \big  [  S^\mu q^\nu   - S^\nu   q^\mu \big ]   
+   {  1 \over m_{\cal N}  } (  p_1^\mu  q^\nu  -p_1^\nu   q^\mu   ) {\bf 1}^{2J}
 \bigg \}   ( \chi_3 \sigma^\mu  \bar \sigma^\nu  \lambda_4 )    
. \end{align}

Below we evaluate the Coulomb corrections proportional to $C_S^i$ and  $C_T^i$ up to next-to-leading order in $1/m_{\cal N}$ expansion. 
A complete analysis at that order should also include non-SM corrections to the next-to-leading beta decay amplitude in \cref{eq:BDA_Mbeta1}. This would involve a large number of new Lorentz structures with corresponding Wilson coefficients (see Ref.~\cite{Falkowski:2021vdg} for a partial list), and therefore the task is left for a future publication.   

The calculation of Coulomb corrections in the presence of the scalar and tensor interactions in \cref{eq:beta_bsm} proceeds in complete analogy to the SM case considered in \cref{sec:CC}. 
Therefore we only list the final results for the discontinuity.
We use the notation in \cref{eq:CC_Master_Discij} and split the discontinuity as 
$D^{i,j} = \sum_{X=V,A,S,T} C_X^i D_X^{i,j}$. 
The vector and axial parts computed in \cref{sec:CC} remain unchanged. 
For the scalar and tensor parts we find 
\begin{align}
    D_S^{0,0}    = &   
i { M_F    q_e Z' e^2   \over 4 \pi \mathcal{K}_e  } \bigg\{   m_e^2 (\chi_3 \lambda_4) 
 +  m_e \bigg[1 + \frac{(p_1 p_3)}{m_\N^2}\bigg] (p_1^\mu - p_2^\mu) 
 (\tilde \chi_3 \bar \sigma_\mu   \lambda_4 ) \bigg\}     
 - 2 i \phi  M_F m_\N (\chi_3 \lambda_4)
 , \nnl 
 D_T^{0,0}  = & 
 i  {  q_e Z' e^2 M_F     m_e  \over 2 \pi  m_\N \K_e  }    { S^\mu  \over J}
 \bigg \{ 
 \bigg [ 1 -  {(p_1 p_3) \over m_{\cal N}^2 } \bigg ] 
 m_{\cal N}^2  (\tilde \chi_3   \bar \sigma_\mu  \lambda_4 )  
         -  2 p_4^\mu (\tilde \chi_3 p_1 \bar \sigma   \lambda_4 )  
 \bigg \} 
 \nnl &    
 - i \phi  
 { 2 M_F S^\mu  \over J}( \chi_3  p_1  \sigma   \bar \sigma_\mu  \lambda_4 ) 
 , \nnl 
D_S^{0,1} = &
\frac{ e q_e M_F \mu_{\cal N'} p_2^{\nu} S^{\beta} 
    \epsilon_{\mu \nu \alpha \beta}}{4 \pi J m_{\mathcal{N}} \mathcal{K}_e}  
\Bigg\{ 2 p_3^{\alpha} \big[m_e m_{\mathcal{N}}^2 
\big(\tilde{\chi}_3 \bar \sigma^{\mu} \lambda_4\big)
        - \big(p_2 p_3\big) \big(\chi_3 \sigma^{\mu} p_2 \bar{\sigma} \lambda_4\big)\big]
    \notag\\
    &\quad + m_{\mathcal{N}}^2 \mathcal{K}_e^2 \bigg[
        \big(c_2 p_2^{\alpha} + c_3 p_3^{\alpha} + c_4 p_4^{\alpha}\big) 
        \big[c_3 \big(m_e \big(\tilde{\chi}_3 \bar \sigma^{\mu} \lambda_4\big) 
        + 2 \big(\chi_3 \lambda_4\big) p_3^{\mu}\big) \notag\\
    &\quad + c_2 \big(\chi_3 \sigma^{\mu} p_2 \bar{\sigma} \lambda_4\big)\big] 
        + c_{\epsilon}^2 {\epsilon}_{ABCD} p_2^B p_3^C p_4^D {\epsilon}^{\alpha}_{\;LMN} p_2^L p_3^M p_4^N 
         \big(\chi_3 \sigma^{\mu} \bar \sigma^A \lambda_4\big) 
         \bigg] \Bigg\}
, \nnl  
D_T^{0,1} = &  
-\frac{e q_e M_F \mu_{\cal N'} p_2^{\nu} S_{\rho} \bar S{}^{\beta} 
    {\epsilon}_{\mu \nu \alpha \beta}}{4 \pi J^2 m_{\mathcal{N}} \K_e}  
\Bigg\{ 
2 p_3^{\alpha} \bigg[
        m_e \big(\tilde{\chi}_3 \bar{\sigma}^{\mu} p_2 \sigma \bar{\sigma}^{\rho} \lambda_4\big)
        - \big(p_2 p_3\big) \big(\chi_3 \sigma^{\mu} \bar\sigma^{\rho} \lambda_4\big)
    \bigg] \notag\\
    &\quad +  \K_e^2 \bigg[
        \big(c_2 p_2^{\alpha} + c_3 p_3^{\alpha} + c_4 p_4^{\alpha}\big) \big(c_2 p_2^{\gamma} + c_3 p_3^{\gamma} + c_4 p_4^{\gamma}\big) 
        \big(\chi_3 \sigma^\mu \bar \sigma_\gamma p_2  \sigma \bar \sigma^\rho \lambda_4\big)  \notag\\
    &\quad +  c_{\epsilon}^2 {\epsilon}^{\alpha}_{\; LMN} p_2^L p_3^M p_4^N {\epsilon}_{\gamma BCD} p_2^B p_3^C p_4^D \big(\chi_3 \sigma^\mu \sigma^\gamma p_2  \sigma \bar \sigma^\rho \lambda_4\big)
    \bigg]
\Bigg\}
. \end{align}
These translate into non-SM contributions to the correlation coefficients, including the $T$-odd ones. 
In the following we focus on the $D$ parameter. 
We get 
\begin{align}
\label{eq:BSM_D-tree}
\Delta_{\rm BSM} D^{\rm tree} \xi_0  = &
\im \bigg \{ 
2 C_S^i  \bar C_T^i   
+  {m_e \over m_{\cal N} } 
\bigg [ 
 \mp   C_T^i \bar C_V^i 
   + {2 \over J}  C_T^i \bar C_{A}^i
   + C_S^i  \bar C_M^i 
\mp  { 1 \over 2 J  } C_T^i \bar C_M^i 
\bigg ] \bigg \} 
+ \cO(m_{\cal N}^{-2})
,\end{align} 
\begin{align}
\label{eq:BSM_D-coulomb}
\Delta_{\rm BSM}  D^C  \xi_0 = & 
-\frac{ e^2 q_e Z' m_e}{8 \pi \K_e } \re \bigg\{
\pm 4   C_V^i \bar C_T^i
\mp 4  C_A^i \bar C_S^i
\notag\\ & 
\mp 4 { (p_1 p_3) \over m_{\cal N}^2 }
C_V^i \bar C_T^i
\mp 4{(p_1 p_3)  \over m_{\cal N}^2 } 
C_A^i \bar C_S^i
+ 2 {(3 p_1 p_3 + p_1 p_4) \over J  m_{\cal N}^2 } C_A^i \bar C_T^i 
\notag\\ & 
-2 {(p_1 p_3 - p_1 p_4)\over m_{\cal N}^2 } C_M^i \bar C_S^i
\pm {(p_1 p_3 + p_1 p_4) \over J m_{\cal N}^2 }  C_M^i \bar C_T^i
\notag\\ &  
+ { (p_1 p_3)^2 + 3 m_e^2 m_{\cal N}^2 
\over 2 m_e m_{\cal N}^3  } C_S^i \bar C_T^i 
-  { 3(p_1 p_3)^2 +  17 m_e^2 m_{\cal N}^2 
\over 4 J m_e m_{\cal N}^3 } |C_T^i|^2 
  \bigg\}
   \notag \\
& \pm \frac{3  q_e e \mu_{\N'} \K_e} 
{8 \pi  J^2 } \bigg \{ 
-J |C_S^i|^2 
\pm \re[  C_S^i \bar C_T^i]  
+ (J+1) |C_T^i|^2 
+ \cO(m_{\cal N}^{-2})
\bigg \}
, \end{align} 
where the normalization factor is now 
$\xi_0 = |C_V^i|^2 + |C_S^i|^2
+ {J+1 \over J} \big ( |C_A^i|^2 + |C_T^i|^2 \big ) $.
The first term in \cref{eq:BSM_D-tree} is the leading order result from Ref.~\cite{Jackson:1957zz}.
The remaining terms are next-to-leading order corrections, first calculated in Ref.~\cite{Falkowski:2021vdg}.
To be non-zero, the tree-level expression requires imaginary parts of the Wilson coefficients, but these are extremely constrained by experiment, especially in the tensor case~\cite{Falkowski:2022ynb}.
In this case, Coulomb corrections allow one to generate non-SM contributions to the $D$ parameter with real $C_S^i$ and $C_T^i$. 
The first term in the curly bracket in  \cref{eq:BSM_D-coulomb} was first computed in Ref.~\cite{Jackson:1957auh}. 
More recently, it was discussed in Ref.~\cite{Falkowski:2022ynb} with the conclusion that the effects of $C_S^i$ and $C_T^i$ may be observable via future precision measurements of $D$.
The remaining terms are subleading Coulomb corrections beyond the SM. 
The scalar contributions  can be read off from Ref.~\cite{Holstein:1972bbl}. 
The tensor contributions are new. 
Given  that $|C_S^i|/C_V^i, |C_T^i|/C_V^i \lesssim 10^{-3}$~\cite{Falkowski:2020pma},  neither of these subleading corrections may be observable in the foreseeable future as their contributions to $D$ are $\cO(10^{-7})$.  

We close this section by noting that, in the presence of the real tensor Wilson coefficient, both $\tilde c_3$ and $\tilde c_4$ are generated  at $\cO(m_{\cal N}^{-1})$ and linear in $C_T^i$. 
This leads to possible $\cO(10^{-7})$ BSM contributions to these $T$-odd correlation coefficients. 

\section{Conclusions} 
\label{sec:conclisions}

In this paper we performed a new study of Coulomb corrections in nuclear beta decay. 
We proposed a novel formulation relying on on-shell techniques and a recently introduced spinor helicity formalism for massive particles. 
This allows one to write down  Lorentz invariant and little group covariant amplitudes for beta decay processes for any spins of the parent and daughter nuclei. 
Focusing on the mixed Fermi--Gamow-Teller allowed decays, we calculated the Coulomb corrections to the amplitude  due to final-state interactions between the outgoing nucleus and beta particle. 
We worked at one loop ($\cO(\alpha)$) and up to  the subleading order  ($\cO(m_{\cal N}^{0})$) in the expansion in the inverse power of nuclear mass.   
The result was cast in a compact Lorentz invariant form. 

Coulomb corrections contribute to the correlation coefficients in the differential distribution of beta decay products.   
Among those,  the most interesting phenomenologically are corrections to the $T$-odd correlations, as they can be generated with real Wilson coefficients. 
We calculated the Coulomb corrections to the $D$ parameter, reproducing the known results in the literature~\cite{Callan:1967zz,Holstein:1972bbl,Jackson:1957auh} in the SM limit, and determining next-to-leading order effects due to scalar and tensor interactions beyond the SM.
We re-evaluated the SM predictions for $D$ for a number of mirror transitions using the most recent experimental input. 
Finally, we calculated Coulomb corrections to the so-called $\tilde c_3$ and $\tilde c_4$ $T$-odd correlations coefficients. The former can be non-zero for transitions with spin $J \geq 1$ and in the SM limit is generated at the same order as the $D$ parameter, while the latter is found to be zero at said order. 

The formalism we propose makes it relatively straightforward to generalize the calculation of Coulomb corrections to higher orders. 
Both higher-loop and subsubleading effects in momentum exchange can be targeted. 
This would allow for a more careful estimation of uncertainties due to higher order SM contributions to $D$, in view of the expected future precision measurements of that correlation. 
It would also be interesting to perform a complete analysis of non-SM effects for Coulomb corrections at the subleading order. 
More generally, the formalism can be employed to facilitate calculations of radiative effects other than the Coulomb corrections. 
Finally, it can be applied to other types of beta transitions, in particular to forbidden decays, to better understand the impact of radiative effects, along the lines of Ref.~\cite{Seng:2024zuc}.

\section*{Acknowledgements}

We would like to thank Ajdin Palavri\'c  and Antonio Rodr\'\i{}guez-S\'anchez for discussions and collaboration at an early stage of this project. 
We are grateful to Mart\'\i{}n Gonz\'alez-Alonso for discussions and comments on the manuscript. 
AF has received funding from the Agence Nationale de la Recherche (ANR)  grant ANR-19-CE31-0012 (project MORA) and from the European Union’s Horizon 2020 research and innovation programme  the Marie Skłodowska-Curie grant agreement No 860881-HIDDeN.

\appendix

\newpage

\section{Useful formulae} 
\label{app:CON}

In this appendix we show the mechanics of the multiplication of spin vectors and we collect a number of useful formulae concerning spin structures and leptonic sum rules.

Recall that the spin vector of a particle of mass $m$ and spin $J$, characterized by  momentum $p^\mu$ and spinors 
$p \sigma = \chi^L \tilde \chi_L$ is defined as 
 \begin{align} 
  [S^\mu]_{O_1 \dots O_{2J} }^{I_1 \dots I_{2J} }  \equiv  &
 - {J \over 2 m } 
 \big [  \chi_{O_1}  \sigma^\mu    \tilde   \chi^{I_1}   +   \tilde  \chi_{O_1}  \bar \sigma^\mu   \chi^{I_1} \big ]  
 \delta_{O_2}^{I_2} \dots  \delta_{O_{2J}}^{I_{2J}}
  \end{align} 
 with upper and lower indices implicitly symmetrized. 
The spin vector encodes the polarizations of the parent and daughter nuclei: the upper little group indices belong to the former, while the lower ones to the latter.
 Complex conjugation acts as
  \begin{align}  
 \bigg (   [S^\mu]_{O_1 \dots O_{2J} }^{I_1 \dots I_{2J} } \bigg )^* 
   = [S^\mu]^{O_1 \dots O_{2J} }_{I_1 \dots I_{2J} } 
  ,  \end{align}  
 thus $S^\mu$  is hermitian. 

In the discontinuity or in the amplitude squared summed over daughter polarization  we may encounter a product of two spin vectors with the structure
$[\bar S^\mu S^\nu]^{I_1 \dots I_{2J} }_{I_1' \dots I_{2J}' } \equiv 
\sum_{O_1 \dots O_{2J} }
[S^\mu]^{O_1 \dots O_{2J} }_{I_1' \dots I_{2J}'} 
[S^\nu]_{O_1 \dots O_{2J} }^{I_1 \dots I_{2J} }$. 
The product can be decomposed as 
        \begin{align} 
\label{eq:spin_multiplication}
  & 
 \bar S^\mu S^\nu  
 =  -  {i     \over 2 m}    \epsilon^{\mu \nu \alpha \beta}  p^\alpha S^\beta 
+ { p^\mu p^\nu  - m^2 \eta^{\mu \nu} \over m^2}  { J (J+1) \over  3} {\bf 1}^{2J} 
+ S^{\mu \nu}
,  \end{align} 
where for $J > 1/2$ we define the transverse and traceless symmetric spin tensor: 
 \begin{align}  
   S^{\mu \nu}  \equiv & 
   \hat S^{\mu \nu} 
   - { p^\mu p^\nu  - m^2 \eta^{\mu \nu} \over m^2}  { J (2J-1) \over  6} {\bf 1}^{2J} 
, \nnl   
\,  [\hat S^{\mu \nu}]_{K_1 \dots K_{2J} }^{L_1 \dots L_{2J} } \equiv & 
 { J (2J-1) \over 8 m^2 }
   \big [  \chi_{K_1}  \sigma^\mu    \tilde   \chi^{L_1}   +   \tilde  \chi_{K_1}  \bar \sigma^\mu   \chi^{L_1} \big ] 
  \big [  \chi_{K_2}  \sigma^\nu    \tilde   \chi^{L_2}   +   \tilde  \chi_{K_2}  \bar \sigma^\nu   \chi^{L_2} \big ]  \delta_{K_3}^{L_3} \dots  \delta_{K_{2J}}^{L_{2J}}
  . \end{align} 
Once again, the upper and lower spin indices are separately symmetrized above.
The spin tensor satisfies 
$p_1^\mu S_{\mu \nu} = p_1^\nu S_{\mu \nu}  
= \eta^{\mu \nu }S_{\mu \nu} = 0$ and 
$S_{\mu \nu} = S_{\nu \mu}$.   
 As a sanity check, note that \cref{eq:spin_multiplication}   implies  
         \begin{align}  
 \bar S^\mu S_\mu   =  & 
 -  J (J+1)  {\bf 1}^{2J} 
  . \end{align}  
Moreover, we also encounter multiplication of spin vector and tensor:
 \begin{align}    
 \bar S^\mu  S^{\nu \rho} =  &  
      { (2J-1)(2J+3) \over 10  m^2 }   \bigg [ 
    {1 \over 2}  \big [ p^\mu p^\nu  - m^2 \eta^{\mu \nu} \big ] S^\rho 
 +     {1 \over 2}  \big [ p^\mu p^\rho  - m^2 \eta^{\mu \rho} \big ] S^\nu  
\notag \\
&- { 1 \over 3}   \big [ p^\rho p^\nu  - m^2 \eta^{\rho \nu} \big ]   S^\mu  
\bigg ]   -  {i \over 2 m} \epsilon^{\mu \nu \alpha \beta}  p^\alpha  S^{\beta \rho} 
    -  {i \over 2 m} \epsilon^{\mu \rho \alpha \beta}  p^\alpha  S^{\beta \nu} 
    +  S^{\mu\nu \rho} 
,   \end{align} 
   where  for $J>1$ we define 
\begin{align} 
 [S^{\mu\nu \rho}]_{K_1 \dots K_{2J} }^{L_1 \dots L_{2J} } \equiv &  
          - {J (2J-1)(J-1) \over 16  m^3  }
 \big [  \chi_{K_1}  \sigma^\mu    \tilde   \chi^{L_1}   +   \tilde  \chi_{K_1}  \bar \sigma^\mu   \chi^{L_1} \big ] 
  \big [  \chi_{K_2}  \sigma^\nu    \tilde   \chi^{L_2}   +   \tilde  \chi_{K_2}  \bar \sigma^\nu   \chi^{L_2} \big ]  
   \notag\\
   &\big [  \chi_{K_3}  \sigma^\rho   \tilde   \chi^{L_3}   +   \tilde  \chi_{K_3}  \bar \sigma^\rho   \chi^{L_3} \big ]  
\delta_{K_4}^{L_4} \dots  \delta_{K_{2J}}^{L_{2J}}  
-  { (2J-1)(J-1) \over 10 m^2} \bigg [    
    \big [ p^\mu p^\nu  - m^2 \eta^{\mu \nu} \big ]   S^\rho \notag\\
    &    +    \big [ p^\mu p^\rho  - m^2 \eta^{\mu \rho} \big ]   S^\nu
    +    \big [ p^\rho p^\nu  - m^2 \eta^{\rho \nu} \big ]   S^\mu \bigg ]_{K_1 \dots K_{2J} }^{L_1 \dots L_{2J} }
.   \end{align} 
This is transverse, symmetric in all indices,  and traceless, analogously as for the two-index tensor.
Higher orders in the expansion in $1/m_{\cal N}$  would require 
introducing spin tensors with more indices (if $J$ is large enough to support them).  
Notice that, at the order we are working, the 3-index spin tensor (as well as higher order ones) does not  affect the calculation of the correlation coefficients we consider ($D$ and $\tilde c_{3,4}$). This is because our expressions for the Coulomb corrections involve at most two spin vectors.
Incidentally, the 3-index spin tensor also cancels out in the amplitude squared, thus no new correlation coefficients arise at the said order. 
 
The other ingredient we need in order to compute correlation coefficients are sum rules for leptonic currents. These are generally given by
\begin{align}&  
   ( \chi_{3J} \sigma^{\mu_1}\ldots\bar \sigma^{\mu_n} \lambda_4 )  
    (  \chi_{3J} \sigma^{\nu_1}\ldots\bar \sigma^{\nu_m} \lambda_4 )^* =  
 {\rm Tr}[ p_4 \sigma \bar \sigma^{\nu_m}\ldots \sigma^{\nu_1} p_3 \bar \sigma \sigma^{\mu_1}\ldots \bar\sigma^{\mu_n}    ]
, \end{align} 
    \begin{align}&  
   (\tilde \chi_{3J} \bar\sigma^{\mu_1}\ldots\bar \sigma^{\mu_n} \lambda_4 )  
    (  \chi_{3J} \sigma^{\nu_1}\ldots\bar \sigma^{\nu_m} \lambda_4 )^* = -m_e 
 {\rm Tr}[ p_4 \sigma \bar \sigma^{\nu_m}\ldots \sigma^{\nu_1} \bar \sigma^{\mu_1}\ldots \bar\sigma^{\mu_n}    ]
, \end{align} 
    \begin{align}&  
   (\tilde \chi_{3J} \bar\sigma^{\mu_1}\ldots\bar \sigma^{\mu_n} \lambda_4 )  
    ( \tilde \chi_{3J} \bar\sigma^{\nu_1}\ldots\bar \sigma^{\nu_m} \lambda_4 )^* = 
 {\rm Tr}[ p_4 \sigma \bar \sigma^{\nu_m}\ldots \bar\sigma^{\nu_1} p_3  \sigma \bar \sigma^{\mu_1}\ldots \bar\sigma^{\mu_n}    ]
, \end{align} 
 where a trace of a generic number of sigma matrices is found iterating 
 \begin{align}
     &\sigma^\mu \bar \sigma^\nu \sigma^\alpha = \eta^{\mu \nu} \sigma^\alpha + \eta^{\nu \alpha} \sigma^\mu - \eta^{\mu \alpha} \sigma^\nu + i \epsilon^{\mu \nu \alpha \beta} \sigma_\beta, \\
     &\bar \sigma^\mu \sigma^\nu \bar \sigma^\alpha = \eta^{\mu \nu} \bar \sigma^\alpha + \eta^{\nu \alpha} \bar \sigma^\mu - \eta^{\mu \alpha} \bar \sigma^\nu - i \epsilon^{\mu \nu \alpha \beta} \bar \sigma_\beta ,
 \end{align}
 until the only structure remaining is $ {\rm Tr}[\sigma^\mu \bar \sigma^\nu] =2 \eta^{\mu\nu} $. In this work we needed sum rules with up to $n=3,m=2$; higher orders in the nucleon mass expansion will involve higher values of $n,m$ as well.
%

\section{Cut Integrals} 
\label{app:INT}

We parametrize the internal momenta entering in \cref{eq:CC_Master_Disc} as
 \begin{align} 
p_X = &  
 - \alpha  p_2 +   \bigg [ \alpha {(p_2 p_3) + m_\N^2  \over (p_2 p_3) + m_e^2 }   - 1\bigg ] p_3  
 +   \K_e  
 \bigg [\frac{\alpha}{\alpha_0} \bigg(1  -  {\alpha \over \alpha_0 }\bigg)  \bigg ]^{1/2} \big [  z   w +   z^{-1} \bar w \big ] 
, \nnl 
p_Y = & 
 - ( 1 - \alpha )  p_2 -  \alpha {(p_2 p_3) + m_\N^2  \over (p_2 p_3) + m_e^2 }  p_3 
 -   \K_e  
 \bigg [\frac{\alpha}{\alpha_0} \bigg(1  -  {\alpha \over \alpha_0 }\bigg)  \bigg ]^{1/2} \big [  z   w +   z^{-1} \bar w \big ] 
. \end{align} 
where $p_2^2 =  p_Y^2 = m_{\cal N}^2$,  $p_3^2 = p_X^2 =  m_e^2$, $\alpha \in [0,\alpha_0], \alpha_0 = \frac{2 (p_2 p_3) + m_e^2}{2 (p_2 p_3) +m_\N^2 + m_e^2 }$, and $z\in \mathbb{C}$ is integrated along the unit circle centered around the origin. 
The Lorentz vector $w^\mu$ is defined via the constraints 
$p_2 w = p_2 \bar w =p_3 w = p_3 \bar w = w^2 = \bar w^2 =0$, which are solved using the ansatz
\begin{align} 
  w^\mu = & 
  c_2 p_2^\mu + c_3 p_3^\mu+ c_4 p_4^\mu 
  + i c_\epsilon  \epsilon^{\mu \nu \alpha \beta}p_2^\nu p_3^\alpha p_4^\beta 
 ,\nnl   
 \bar  w^\mu = & 
 c_2 p_2^\mu + c_3 p_3^\mu+ c_4 p_4^\mu  
 -  i c_\epsilon \epsilon^{\mu \nu \alpha \beta}p_2^\nu p_3^\alpha p_4^\beta 
.\end{align}
This leads to the conditions
\begin{align} 
0 = & c_2 m_{\cal N}^2  + c_3 (p_2 p_3) + c_4 (p_2 p_4)      
,\nnl 
0 = & c_2(p_2 p_3)  + c_3 m_e^2  + c_4 (p_3 p_4) 
,\nnl  
-1 =  & 
( c_2 p_2^\mu + c_3 p_3^\mu+ c_4 p_4^\mu )^2 
, \nnl 
-1 =  &   (c_\epsilon \epsilon^{\mu \nu \alpha \beta}p_2^\nu p_3^\alpha p_4^\beta)^2  
\end{align} 
solved by
 \begin{align}  
 \label{eq:DIST_c234e}
   c_2 = & 
 {(p_2 p_3)(p_3 p_4) 
 - m_e^2 (p_2 p_4) \over 
 m_\N \K_e 
\sqrt{ 2 (p_2 p_3) (p_2 p_4) (p_3 p_4)  
-  m_{\cal N}^2 (p_3 p_4)^2  
- m_e^2 (p_2 p_4)^2 }     }  
   , \nnl
 c_3 = & 
 {(p_2 p_3)(p_2 p_4) - m_{\cal N}^2 (p_3 p_4) \over 
 m_\N \K_e 
\sqrt{ 2 (p_2 p_3) (p_2 p_4) (p_3 p_4)  
-  m_{\cal N}^2 (p_3 p_4)^2  
- m_e^2 (p_2 p_4)^2 }     }  
      , \nnl
c_4 = & 
{m_{\cal N}^2m_e^2 - (p_2 p_3)^2   \over 
m_\N \K_e  
\sqrt{ 2 (p_2 p_3) (p_2 p_4) (p_3 p_4)   -  m_{\cal N}^2 (p_3 p_4)^2  - m_e^2 (p_2 p_4)^2 }     }    
         , \nnl 
 c_\epsilon=  & 
  {1 \over  
\sqrt{ 2 (p_2 p_3) (p_2 p_4) (p_3 p_4)   -  m_{\cal N}^2 (p_3 p_4)^2  - m_e^2 (p_2 p_4)^2 }     } 
 . \end{align}
For power counting, it is useful to remember that $c_2 \sim c_\epsilon \sim \cO(m_{\cal N}^{-1})$, while 
  $c_3 \sim c_4 \sim \cO(m_{\cal N}^{0})$.
The spurious poles present in $c_i$ corresponding to 
$2 (p_2 p_3) (p_2 p_4) (p_3 p_4)   -  m_{\cal N}^2 (p_3 p_4)^2  - m_e^2 (p_2 p_4)^2  \to 0$ are artifacts of our parametrization and cancel out in observables. 

Using this parametrization, the discontinuities considered throughout out work always reduce to the ones of the following two integrals:
 \begin{align}  
&I_{\circ} \equiv  \int {d^4 k \over i (2 \pi)^4}
{1 \over 
\big [  k^2 - m_Y^2  + i \epsilon \big ]  \big [  (k + p_2 + p_3)^2 - m_X^2  + i \epsilon  \big ]   },\\
&I_{\triangle} \equiv  \int {d^4 k \over i (2 \pi)^4}
{1 \over 
\big [  k^2 - m_Y^2  + i \epsilon \big ] \big  [ (k+p_2)^2 + i \epsilon \big ]
 \big [  (k + p_2 + p_3)^2 - m_X^2  + i \epsilon  \big ]   }
.  \end{align} 
These are readily computed as 
\begin{align}  
&{\rm Disc}_2^u I_{\circ}  =  - \int  \text{d}\Pi_{XY} =  - {  \K_e   \over 4 \pi m_\N}, \\
&{\rm Disc}_2^u I_{\triangle}  = - \int  { \text{d}\Pi_{XY} \over (p_X + p_3)^2 } = 
{1 \over 8 \pi  u \beta_u} 
\int_0^{\alpha_0} {d \alpha \over \alpha}
. \end{align}  
Notice that the triangle discontinuity is IR divergent as $\alpha$ approaches zero, as expected.
The IR divergence is controlled by soft theorems and cancels out in correlation coefficients.

\bibliographystyle{JHEP}
\bibliography{refs} 

\end{document}